\documentclass[a4paper,12pt]{article}
\usepackage[cp1251]{inputenc}
\usepackage[T2A]{fontenc}
\usepackage{longtable}
\usepackage{latexsym}
\usepackage[dvips]{graphicx}
\usepackage{amssymb}
\usepackage{amsfonts}
\usepackage{amsmath}
\usepackage[english]{babel}

\topmargin = - 1cm
\hoffset = - 1 cm

\begin{document}

MSC 85A25, 81P20
\vskip 0.3cm

\begin{center}
{\bf Energy flux of electromagnetic field in stochastic model\\
of radiative heat transfer in dielectric solid medium}
\medskip

$\boldsymbol{Yu.P.\ Virchenko\ \ and\ \ Lam\ Tan\ Phat}$\\
{\small \it Belgorod State University, 308015, Belgorod, Russia}
\end{center}

\newcommand{\avel}{\langle \hskip - 0.25cm\langle\,}
\newcommand{\aver}{\,\rangle \hskip - 0.25cm\rangle}

{\small The stochastic model that  describes  radiative  heat   transfer in dielectric medium is built. The model is based on the representation that heat transfer is realized both by heat conductivity mechanism in it and due to the electromagnetic radiation that is generated by thermal fluctuations of atoms in the medium. Using the fluctuation-dissipative theorem, on the basis of such  physical suppositions,  the stochastic model is formulated in the form of the infinite dimensional Ornstein-Uhlenbeck process that describes medium fluctuations. In the model frameworks, the energy flux density of fluctuating electromagnetic field is calculated in the form  of the functional of  temperature distribution in three-dimensional medium sample.}
\medskip

{\bf 1. Introduction.} The heat transfer in solids is realized by   two mechanisms. They are the proper thermal conductivity and the heat transfer by electromagnetic radiation. The last is generated by thermal fluctuations of the medium local thermodynamic state. In accordance with such physical representation, the evolution equation of the temperature distribution $T({\bf x}, t)$ may be written phenomenologically in the form  at each fixed time  $t$ (see, for example, \cite{Spar}-\cite{Petr})
$$\kappa (T) {\dot T}({\bf x}, t) = (\nabla, \varkappa(T) \nabla) T({\bf x}, t) - (\nabla, {\bf S})({\bf x}, t)  \eqno (1)$$
where $\varkappa (T)> 0$ is  the medium thermal conductivity coefficient which depends on temperature, $\kappa (T)$ is the medium volume heat capacity. The vector field ${\bf S}({\bf x}, t)$ is the energy flux density of electromagnetic radiation associated with those absorption and radiation actions of electro\-mag\-ne\-tic field by means of which the heat is  transferred.

The $(\nabla, {\bf S}({\bf x}, t))$ value  multiplied on the small volume of the spatial medium region centered near the point ${\bf x}$ is the part of flux density  which is spent in the medium heating its small volume centered in the point ${\bf x}$ at the time moment $t$. This term in Eq.~(1) is essential  when heat transfer problems are solved in optically semitransparent medium that possesses a low electrical conductivity and also some sufficiently large temperature drops are present through a
characteristic distance ${\bar L}$. To solve the problems of heat transfer in these cases, we must obtain a complete evolution equation controlling the temperature distribution $T({\bf x}, t)$. So, it is necessary to find the explicit form of the functional ${\bf S}({\bf x}, t) = {\bf S}[T({\bf x}, t)]$ which transforms Eq.~(1) into the self-consistent one.

Usually, the energy flux density ${\bf S}({\bf x}, t)$ is constructed phenomenologically in frameworks of so-called theory of radiation transfer. It is done using: the  geometric optics laws which are applied to <<thermal>> rays inside the medium, the phenomenological Kirchhoff law that concerns the  radiation and absorption intensities of the optic radiation, as well as the Beer-Bouguer-Lambert law (see, for example, \cite{Spar}-\cite{Kol}). The thermal electromagnetic field in such theoretical constructions does not exist in such a theory. It seems that such a situation is unsatisfactory from the theoretical viewpoint. It is connected with the absence of successive microscopic theory of radiation heat transfer which should be based on the quantum theory of radiation and absorption of thermal photons in solid medium.

Here, we shall not concentrate on detailed analysis of those problems which are related with the construction of the microscopic theory of heat radiation transfer based on  statistical physics formalism. We point out only  that the statistical approach in the theory of heat radiation transfer has been proposed in the Rytov works which are summarized in the monographs \cite{Ryt1}, \cite{Ryt2}. In connection with the complexity of microscopic theory construction, this approach is semi-phenomenological. It is based on the presentation that the electro\-mag\-ne\-tic  radiation and absorption actions in the medium are connected  with thermal fluctuations of its local thermodynamic state. These thermal fluctuations determine the microscopic fluctuations of charges in medium and currents induced by them. Such a fluctuation approach is the statistical one due to its nature. Therefore, the electromagnetic field which is responsible for the radiation heat transfer in the medium, is the stochastic one. Besides, the microscopic mechanism  of the energy field  transformation into heat is not concretized in the framework of such a theory. It permits to avoid the quantum description of the radiation and absorption actions. It turn, it is reasonable from theoretical viewpoint, since the heat radiation transfer is not a quantum effect.

We note that thermal fluctuations of electric charges which generate the stochastic electromagnetic field, may be occurred in electro-neutral mediums having very low
electrical conductivity, i.e. in dielectrics and high-resistance semiconductors. Thus, thermal fluctuations of charges lead  with inevitability to induction of electric currents in such media. But these currents  exist for very small distances. Since the amplitude of thermal fluctuations increases with the temperature growth, then, for sufficiently large its value, thermal vibrations of medium atoms lead to electric charges fluctuations even in dielectric media.  They are occurred for distances having the order of interatomic ones. Their value may be appeared essential when the problem of heat transfer caused by the electromagnetic radiation is solved.

The mathematical realization of the described physical considerations is performed by the use of stochastic electromagnetic fields which are obeyed Maxwell's equations.
At present work, we construct the specific mathematical stochastic model of thermal radiation transfer in frameworks of the above-described fluctuation approach.
To avoid the  account of the boundary conditions in the suggested model, we study only the case when the inhomogeneity of temperature distribution is concentrated in a limited region of boundless medium envi\-ron\-ment.
\smallskip

{\bf 2. The mathematical model construction.} For the problem formulation, we consider that  the thermal electromagnetic field is defined by the field pair $\langle {\tilde {\bf E}} ({\bf x}, t), {\tilde {\bf H}}({\bf x}, t)\rangle$, ${\bf x} \in {\Bbb R}^3$, $t \in {\Bbb R}$, which are stochastic ones. Here and after we mark all  random functions by the sign <<tilde>>. On the basis of this pair, the energy flux density of thermal electromagnetic field  is determined as
$${\tilde {\bf S}} ({\bf x}, t)  = \frac c {4 \pi}[{\tilde {\bf E}}, {\tilde {\bf H}}] ({\bf x}, t) \eqno (2)$$
where $c$ is the vacuum light velocity. So, it is a random function.

The thermal electromagnetic field changes rapidly through distances having the charac\-te\-ris\-tic wavelength ($\sim 10^{-4}$~cm) order that corresponds to thermal (red and infra-red) radiation.  Therefore, the characteristic time has the order of $\sim 10^{-14}$~sec. At the same time, the characteristic length of thermal conductivity process in crystalline dielectrics has the order $10^{-2}$~cm, and the correspondent characteristic time is $10^{-1}$~sec. Therefore, the energy flux density (2) should be averaged over spatial regions having a size which is much greater than the characteristic wavelength of stochastic electromagnetic field when adequate mathematical theory is constructed. But it is much smaller than the characteristic length of the heat transfer process. In addition, the density (2) should be averaged over temporal  intervals which are much greater than the characteristic period of thermal radiation oscillations, but it is much smaller than the characteristic time of thermal conductivity process. Such an averaging permits to ignore the small rapid oscillations of the divergence $(\nabla, {\tilde {\bf S}}({\bf x}, t))$ of the radiation flux density over space and time because they have no a relation to the heat transfer process. Due to basic statistical physics representations, the pointed out space-time averaging is equivalent to the averaging on the basis of the probability distribution of random electromagnetic field, when the pair of  random fields ${\tilde {\bf E}} ({\bf x}, t)$  and ${\tilde {\bf H}} ({\bf x}, t)$ possesses the {\it ergodicity} property. Thus, the energy flux density of the field used in (1) is determined by the mathematical expectation ${\bf S}({\bf x}, t) = \avel\, {\tilde  {\bf S}}({\bf x}, t)\aver$ on the probability distribution   of the random vector field ${\tilde  {\bf S}}({\bf x}, t)$ (here and after angular brackets denote such an averaging). Then, for the complete mathematical formulation of the model that describes the radiation heat transfer, it is necessary to build the adequate stochastic model of the thermal electromagnetic field and to calculate the mathematical expectation $\avel {\bf S}({\bf x}, t)\aver$ on basis of its probability distribution.

Thus, the stochastic electromagnetic field is represented  by random realizations $\langle {\tilde {\bf E}} ({\bf x}, t)$, ${\tilde {\bf H}} ({\bf x}, t)\rangle$ which satisfy the system of stochastic Maxwell equations in the {\it continuous dielectric medium} neglecting its dispersion
$$\begin{array}{ll}\displaystyle \frac {\varepsilon}{c}\,\frac {\partial {\tilde {\bf E}}}{\partial t}\ + \frac {4\pi}{c}{\tilde {\bf j}} = [\nabla, {\tilde {\bf H}}]\,, \qquad &  \displaystyle (\nabla, {\tilde {\bf E}}) = \frac {4\pi} \varepsilon\, {\tilde \rho}\,, \vspace{0.4cm} \\ \displaystyle
\frac {\mu}{c}\,\frac {\partial {\tilde {\bf H}}}{\partial t}\ = -[\nabla, {\bf {\tilde E}}]\,, \qquad  & (\nabla, {\tilde {\bf H}}) = 0\end{array}\ \eqno (3) $$
where ${\tilde {\bf E}}$ and ${\tilde {\bf H}}$ are intensities of  electric and magnetic fields of thermal radiation which are generated by heated medium. At the same
time, $\varepsilon$ is the electric permeability of uniform dielectric medium and $\mu$ is the magnetic one. We consider them to be independent on ${\bf x}$ and~$t$.

Generally, the $\varepsilon$ and $\mu$ values depend on the temperature. These dependencies may be substantial at large temperature drops through distances having the characteristic size order that is connected with temperature non-uniformity in the medium. Temperature values in depen\-den\-cies of $\varepsilon$ and $\mu$ on $T$ should be equal to the local temperature $T ({\bf x}, t)$ in the medium. In general case, spatial and temporal derivatives of $\varepsilon (T ({\bf x}, t))$ and $\mu (T({\bf x}, t))$ should be appeared in the Maxwell equations when these dependencies are taken into account. However, these derivatives are extremely small in comparison with those length and temporal scales which are characteristic of thermal radiation due to slowness of dependencies pointed out. Therefore, these derivatives are not taken into account in Eqs.(3).

Random realizations of ${\tilde {\bf E}} ({\bf x}, t)$  and ${\tilde {\bf H}} ({\bf x}, t)$ are determined by stochastic sources ${\tilde {\bf j}}$, ${\tilde \rho}$ since they  are some solutions of the system (1). These sources are some fluctuations of the electric current density and the charge density. They are occurred at  micro-regions having the order of the characteristic wavelength due to thermal fluctuations.

Besides, for complete determination of solutions,  it is important to propose definite initial and boundary conditions corresponding to described physical situation.
As for  boundary conditions,  we shall  study the simplest physical situation when the thermal localized non-uniformity takes place  in unbounded medium. This nonuniformity is concentrated in a bounded region of space with the linear size $L$ having the order of 1cm $\div 10^2$~cm. Then the local medium temperature $T({\bf x}, t)$ tends to a constant when $|{\bf x}|$ tends to infinity.  As for the spatially distributed stochastic sources which are performed by the densities ${\tilde {\bf j}}$, ${\tilde \rho}$ in Eqs.~(3), their specific form  determines completely the constructed model. The consistency condition of the system Eqs.~(3) leads to the fact that these densities satisfy the {\it continuity equation}
$${\dot {\tilde \rho}} + (\nabla, {\tilde {\bf j}}) = 0\,. \eqno (4)$$
Due to such a relation, it is sufficient to determine only the random field  ${\tilde {\bf j}}({\bf x}, t)$ for complete mathematical building of the model.

In our model,  the current density ${\tilde {\bf j}}$ is composed of two parts. The first is the proper stochastic source of electromagnetic field. It is plays the role of  an internal <<electromotive force>> in the medium. It arises as a result of the thermal fluctuations. The second is  determined by Ohm's law $\sigma {\tilde {\bf E}}$. We note that the coefficient $\sigma > 0$. It plays the role of the electrical conductivity. But it is not the genuine macroscopic electrical conductivity of the medium that may be very small in physical situation under consideration. It performs an <<effective electrical conductivity>>  which should be different from zero due to the so-called  {\it fluctuation-dissipative theorem} (see, for example, \cite{Ryt2}). It is necessary to take into account from the mathematical viewpoint in order that a regular dissipative constituent should be in the system of stochastic evolution equations~(1) with additive noise. In turn, it is connected with presence of stationary evolution regime.

With probability one, the part of the fluctuation current density $a({\bf x}, t; T ){\tilde {\boldsymbol\varphi}}$  that serves the stochastic source of electromagnetic field, should be certainly contained the vortical term (a fluctuation <<Foucault current>>) in spite of the radiation transfer occurs in dielectrics (or high-resistance semiconductors). Here, the source intensity  $a({\bf x}, t; T)$ depends functionally on the local temperature $T = T ({\bf x}, t)$. Therefore,  it may be varied spatially and temporally. This varying is much slower in comparison with the change of the thermal electromagnetic field. The irradiation of electromagnetic waves which transfer the heat  is  associated with the availability of the vortical part. In connection with dielectric character of the medium, the fluctuation current (its correlation function) is concentrated at small space scale that has the order of  $10 \div 30$ interatomic distance. Thus, the current density ${\tilde {\bf j}}$ should be replaced in Eqs.~(3) and (4) by ${\tilde {\bf j}}({\bf x}, t) =  {\tilde {\boldsymbol{\varphi}}}({\bf x}, t) a({\bf x}, t; T) + \sigma {\tilde {\bf E}}({\bf x}, t)$  where the intensity $a({\bf x}, t; T )$ should be defined on the basis of statistical physical consideration  for completion of the model construction. We suppose that the squared intensity is determined by thermal photons irradiation in a small spatial region which concentrates near the point  ${\bf x}$ at the time  $t$. Therefore,
$$a^2 ({\bf x}, t; T) = \hbar \int\limits^\infty_{- \infty} \omega^3 f \Big( \frac {\hbar \omega}{{\rm k}\,T({\bf x}, t)}\Big)d \omega \eqno (5) $$
where $f$ is the energy distribution function of irradiated  photons. It depends  on  the tempera\-ture $T({\bf x}, t)$ distribution. Then we obtain that  $a^2 ({\bf x}, t; T ) \sim T^4 ({\bf x}, t)$, when $f$ is the Planck function.

Substitution of the explicit form of ${\tilde {\bf j}}({\bf x}, t)$ into the Eqs.~(3) leads to the  stochastic equations system with the additive noise ${\tilde {\boldsymbol\varphi}}$ where the field ${\tilde {\bf E}}({\bf x}, t)$ is determined by the equation
$$\frac {\partial {\tilde {\bf E}}}{\partial t}\ + \gamma {\tilde {\bf E}} + \frac {4\pi}{\varepsilon}\, a {\tilde {\boldsymbol\varphi}}  = \frac c\varepsilon [\nabla, {\tilde {\bf H}}]\,, \quad \gamma = \frac {4\pi \sigma}{\varepsilon}\,.\eqno (6)$$
Besides, the  evolution equation of the charge density is valid
$${\dot {\tilde \rho}} + \gamma {\tilde \rho} + (\nabla, a {\tilde {\boldsymbol\varphi}}) = 0 \eqno (7)$$
where, as above, we  have neglected spatial derivatives of the temperature distribution. In general case, the coefficient $\sigma$  depends on the local temperature which changes slowly on ${\bf x}$ and $t$. But we neglect this dependence for reasons above pointed out.

The random field ${\tilde {\boldsymbol\varphi}}$ in Eqs.(6),(7) is Gaussian with the zero average value $\avel {\tilde {\boldsymbol\varphi}}({\bf x}, t)\aver = 0$ due to supposed physical smallness of thermal fluctuations. At the same time, we suppose that $\avel {\tilde \rho} ({\bf x}, t)\aver = 0$.  Then the Gaussian field ${\tilde {\boldsymbol\varphi}}({\bf x}, t)$ is completely determined by the pair correlation function  $K_{j_1j_2}({\bf x}_1, t_1;  {\bf x}_2, t_2)=$ $ \avel {\tilde \varphi}_{j_1} ({\bf x}_1, t_1) {\tilde \varphi}_{j_2} ({\bf x}_2, t_2) \aver$. Due to physical reasons, the random field  ${\tilde {\boldsymbol\varphi}}({\bf x}, t)$  is  translationally invariant on ${\bf x}$ in the stochastical sense  and it is stationary on $t$  in the sense of the theory random processes.
Besides, we assume that this field is stochastically isotropic and temporally reversible.  So, its correlation function is represented in the form
$$K_{j_1j_2}({\bf x}_1, t_1;  {\bf x}_2, t_2)= K (|{\bf x}_1 - {\bf x}_2|, |t_1 - t_2|)\delta_{j_1, j_2}\,. \eqno (8)$$
In this case, the source $a({\bf x}, t; T){\tilde {\boldsymbol\varphi}}({\bf x}, t)$  of thermal radiation is uniform on ${\bf x}$ in Eqs.~(6),(7) if we neglect the pointed out slow dependence on the local temperature $T ({\bf x}, t)$. Moreover, at such conditions, it is stationary on $t$ and it is stochastically isotropic.

Further, we use some supplement assumptions about properties of the function $K(r, s)$,  $r, s > 0$. These properties are associated with the locality of correlation function \linebreak $K_{j_1j_2}({\bf x}_1, t_1;  {\bf x}_2, t_2)$.  For physical consideration,  the random field ${\tilde {\boldsymbol\varphi}}({\bf x}, t)$ should have the extreme small correlation time. Such correlations should be disappear during the temporal interval equal to several periods of stochastic electromagnetic field oscillations.  Then, we suppose that $K(r, s) \sim \delta (s)$.  In this case, the field ${\tilde {\boldsymbol\varphi}}({\bf x}, t)$ is transformed to a generalized random Gaussian field of the <<white noise>> type on the temporal variable.  Spatial correlations   of the field ${\tilde {\boldsymbol\varphi}}({\bf x}, t)$ values  are also short-ranged. They disappear at the distance equal to some interatomic lengthes. So, the correlation length is the smallest parameter between those which have the linear size  in the problem under study. However, for the reasons that will become clear from the subsequent analysis, we may not assume that the function  $K(r,s)$  is proportional to $\delta (r)$ by the analogy with the temporal variable. So, we use the next representation
$$K (|{\bf x}_1 - {\bf x}_2|, |t_1 - t_2|) = K(|{\bf x}_1 - {\bf x}_2|)\delta (t_1 - t_2) \eqno (9)$$
where the function $K(r)$  is absolutely integrable $\int_{{\Bbb R}^3} |K({\bf x})|d {\bf x} < \infty$  and it is localized in the zero neighborhood having the $r_0 > 0$ size order that is $K(r) = r^{-3}_0 Q (r^2/2r^2_0)$ where  $r_0$ is a small parameter and $K = \int_0^\infty Q(\xi^2/2) d \xi < \infty$. Here the function $Q(r)$ is concentrated in the region with the linear size of order 1.

After determination of the random process ${\tilde {\bf j}} ({\bf x}, t)$ in the stochastic differential equations system (3), the fluctuation electromagnetic field is completely defined by the requirement of its temporal stationarity. At the same time, the random function ${\tilde S}({\bf x}, t)$  is a functional on  $T({\bf x}, t)$, and its  mathematical expectation
$$ \avel {\tilde {\bf S}} ({\bf x}, t)\aver  = \frac c {4 \pi}\,\avel [{\tilde {\bf E}}, {\tilde {\bf H}}] ({\bf x}, t)\aver \eqno (10)$$
is determined by the  probability distribution of the fluctuation field ${\tilde {\boldsymbol\varphi}}$.
\smallskip

{\bf 3. Small parameters of mathematical model.} Consistent mathematical analysis of the random process which is determined  by the constructed mathematical model of radiation heat transfer is very complicated in the physical situation under consideration. In particular, the averaging in the resulting formula for the energy flux density $S_j ({\bf x}, t)$ of the fluctuation electromagnetic field leads to the complicated expressions which are uncomfortable for its practical application when we solve heat transfer problems of electromagnetic radiation in semi-transparent medium. The significant simplification of these expressions is reached, when the specific physical conditions are taken into account where these transport processes occur. It leads to the detection of  sequence small parameters in the problem under study. Then the natural setting of the mathematical problem consist of the calculation of the $S_j ({\bf x}, t)$ expression in the form of main asymptotic term  when these small parameters tend to zero.

Let ${\bar L}$ be the size of temperature non-uniformity that equals to the linear size of the region where the non-uniform distribution temperature $T({\bf x}, 0)$ varies  at one degree. We note that the characteristic time during which the temperature distribution changing is occurred due to the heat conductivity process, is significantly more than the time ${\bar L}/{\bar c}$ during which the thermal electromagnetic radiation overcomes the distance ${\bar L}$ and goes out of the non-uniformity region (it occurs during $\sim 3\cdot 10^{-13}$sec when ${\bar L} \sim 10^{-2}$ cm) where the heat transfer processes occurs. Therefore, this part of  radiation does not effect on the heat transfer process when it comes out of the system. The natural time for the heat transfer process is determined by the value ${\bar L}^2\kappa /\varkappa$ where the ratio $\varkappa/\kappa$  has the order of  $10^{-3}$ cm${}^2/$sec in the typical physical situation in solid high-resistance semiconductor crystal. Consequently, the typical time of the distribution temperature varying in problems under  consideration is equal to $10^{-1}$s. As a result, we obtain the small parameter $\varkappa/  {\bar L} {\bar c} \kappa \ll 1$ having the order of  $3\cdot 10^{-12}$ where  ${\bar c}^2 = c^2/\varepsilon \mu$ and ${\bar c}$ is the light velocity in the medium, ${\bar c}^2 = c^2/\varepsilon \mu$.

Further, we assume that the medium is very semi-transparent. The characteristic distance of the radiation damping  is much larger in it than the introduced size ${\bar L}$.  In this case, if we use typical values of specific electrical conductivity, the parameter $\gamma {\bar L}/{\bar c}$  has the values in the range $3\cdot (10^{-4} \div 10^{-17}) \ll 1$  in  dielectrics where $\gamma =  {4\pi \sigma}/{\varepsilon}$ has the order of $10^6 \div 10^{-7}$  sec${}^{-1}$. For some semiconductors, the parameter  $\gamma {\bar L}/{\bar c}$ varies in the range $4\cdot(10^{-4} \div 10^{5})$.

As mentioned above, there is another natural small parameter which is  the ratio $r_0/{\bar L}$. This ratio is small in view of the fact that $r_0 \sim 10^{-8}$~cm and ${\bar L} \sim 10^{-1}$~cm, so that $r_0/{\bar L} \sim 10^{-7}$. Thus, we conclude that the following relations $\varkappa/{\bar L} {\bar c} \kappa \ll r_0/L$,
$\gamma L/{\bar c} \ll r_0/L$  between the introduced small parameters are fulfilled  in dielectrics.  As we can see from the above estimates,  the parameter $\gamma {\bar L}/{\bar c}$ is not small for semiconductors in general case. Thus, the calculation of the energy flux density of the fluctuation electromagnetic field will be performed in the form of the main asymptotic term when these parameters tend to zero.

In view of the fact that the transition to the limit is realized by several parameters when the asymptotic calculation is done, it is necessary to specify the transition  character. We assume that these transition are understood as repeated ones. In accordance with the their mentioned typical physical values, the  limit transition order  will be realized in the order of their value, i.e. from small ones to large ones. Thus, the transition to the limit of $r_0/{\bar L} \to 0$  will be produced at the final step of calculations. At the same time, for construction of such calculations, it is necessary to explicitly introduce the parameter ${\bar L}$ into the appropriate formulas. Respectively, all  values of length and time dimensionalities in our model are measured by units of the largest spatial size ${\bar L}$ and the biggest temporal duration ${\bar L}^2\kappa/\varkappa$.
\smallskip

{\bf 4. Construction of the random stationary process.} Since the typical time of thermal conductivity process is the largest parameter of the temporal dimensionality in our model, the first step of the mentioned transition to the limit at the asymptotic value $S_j ({\bf x}, t)$ calculation  is the construction of random stationary process on the basis of random process determined by the stochastic Eqs.~(3) with a fixed initial temperature distribution.  With this aim, we introduce the generalized Fourier expansions of the random stochastic realizations of fields ${\tilde {\bf E}} ({\bf x}, t)$ and ${\tilde {\bf H}} ({\bf x}, t)$,
$${\tilde {\bf E}} ({\bf x}, t) = \int\limits_{{\Bbb R}^3}{\tilde {\bar {\bf E}}}({\bf k}, t)\exp[i ({\bf k}, {\bf x})] d{\bf k}\,,\quad {\tilde {\bf H}} ({\bf x}, t) = \int\limits_{{\Bbb R}^3}{\tilde {\bar {\bf H}}}({\bf k}, t)\exp[i ({\bf k}, {\bf x})] d{\bf k}\,. \eqno (11)$$
Here, ${\tilde {\bar {\bf E}}}({\bf k}, t)$  and ${\tilde {\bar {\bf H}}}({\bf k}, t)$  are generalized random fields on ${\bf k}\in {\Bbb R}^3$. We substitute the expansion (11) in  Eqs.~(3), (5), (6). Then, because of their uniqueness determination on the basis of Fourier's expansions, we obtain the finite equations system of the generalized Fourier-images for each ${\bf k}\in {\Bbb R}^3$,
$$\frac {\partial }{\partial t}{\tilde {\bar {\bf E}}}({\bf k}, t)\ + \gamma {\tilde {\bar  {\bf E}}}({\bf k}, t) +  \frac {4 \pi}\varepsilon
{\tilde {\bar {\bf  j}}}({\bf k}, t) = \frac {ic}{\varepsilon} [{\bf k}, {\tilde {\bar  {\bf H}}}({\bf k}, t)]  \,, \eqno  (12)$$
$$\frac {\partial }{\partial t}{\tilde {\bar {\bf H}}}({\bf k}, t)\ = - \frac{ic}\mu[{\bf k}, {\bf {\bar {\tilde E}}}({\bf k}, t)]\,, \quad ({\bf k}, {\tilde {\bar  {\bf E}}}({\bf k}, t)) = - \frac {4\pi i} \varepsilon\, {\tilde {\bar \rho}}({\bf k}, t) \,,  \quad   ({\bf k}, {\tilde  {\bar {\bf H}}}({\bf k}, t)) = 0\,, \eqno (13)$$
$${\dot {\tilde \rho}} ({\bf k}, t) + \gamma {\tilde \rho}({\bf k}, t) + i ({\bf k}, {\tilde {\bar {\bf j}}}
({\bf k}, t)) = 0\,. \eqno (14)$$
Besides, we introduce generalized Fourier images of random realizations corresponding to charge distribution  density,
$${\tilde  \rho} ({\bf x}, t) = \int\limits_{{\Bbb R}^3}{\tilde {\bar  \rho}}({\bf k}, t)\exp[i ({\bf k}, {\bf x})] d{\bf k}\,. \eqno (15)$$
As well,  we introduce generalized Fourier images ${\tilde {\bar {\bf j}}} ({\bf k}, t)$ of  the random field $a({\bf x}, t){\tilde {\boldsymbol\varphi}}({\bf x}, t)$
realizations,
$$a({\bf x}, t; T){\tilde {\boldsymbol\varphi}}({\bf x}, t) = \int\limits_{{\Bbb R}^3}{\tilde {\bar {\bf j}}} ({\bf k}, t)\exp[i ({\bf k}, {\bf x})] d{\bf k}\,. \eqno (16)$$
The fields ${\tilde {\bar {\bf j}}}({\bf k}, t)$, ${\tilde {\bar \rho}}({\bf k}, t)$ are complex-valued Gaussian random ones due to the Gaussian property of the field ${\tilde {\boldsymbol\varphi}}({\bf x}, t)$. They have zero average values $\avel {\tilde {\bar {\bf j}}}({\bf k}, t) \aver = 0$, $\avel {\tilde {\bar \rho}}({\bf k}, t)\aver = 0$.

In view of the reality of the value  $a({\bf x}, t; T){\tilde {\boldsymbol\varphi}}({\bf x}, t)$,  the field ${\tilde {\bar {\bf j}}}({\bf k}, t)$  realizations  has the following property ${\tilde {\bar {\bf j}}}^{\,*}({\bf k}, t) = {\tilde {\bar {\bf j}}}(-{\bf k}, t)$ with the probability one. Namely, it is completely characterized by the correlation function ${\bar K}_{j_1j_2}({\bf k}_1 , t_1; {\bf k}_2, t_2) = \avel {\tilde {\bar \varphi}}_{j_1}({\bf k}_1, t_1) {\tilde {\bar \varphi}}_{j_2}^*({\bf k}_2, t_2)\aver$. This function is positively definite matrix-function on ${\bf k} \in {\Bbb R}^3$ and $t$. Then, it is associated with the correlation function
$${\bar K}_{ll'}({\bf k}, \omega,{\bf k'}, \omega') = \frac 1{(2\pi)^8}\int\limits_{{\Bbb R}^8}\exp\Big[i(\omega' t' -\omega t) + i\Big(({\bf k}', {\bf x}') - ({\bf k}, {\bf x}) \Big) \Big] K_{ll'}({\bf x}, t; {\bf x}', t')d{\bf x}d{\bf x}'dtdt'\,. \eqno (17)$$

Whereas the properties of stochastic uniformity on ${\bf x}$, stationarity on $t$ and isotropy takes place for the field $\tilde {\boldsymbol \varphi} ({\bf x}, t)$, this correlation function has the form
$${\bar K}_{ll'}({\bf k}, \omega,{\bf k'}, \omega') =  \frac 1 {2 \pi}\delta_{ll'}\delta({\bf k} + {\bf k}'){\bar K} ({\bf k})\delta(\omega' + \omega)\,, \eqno (18) $$
$${\bar K} ({\bf k}) = \frac 1{(2\pi)^3}\int\limits_{{\Bbb R}^3}\exp\Big[- i ({\bf k}, {\bf x})\Big]K(|{\bf x}|)d{\bf x}\,. \eqno (19)$$

Since the equation system is finite at each fixed ${\bf k} \in {\Bbb R}^3$, then it is uniquely solvable when the initial conditions of the generalized random ${\tilde {\bar {\bf  E}}} ({\bf k}, t)$, ${\tilde {\bar {\bf  H}}} ({\bf k}, t)$, ${\tilde {\bar \rho}} ({\bf k}, t)$ realizations are given. It means that the stochastic model of thermal electromagnetic field described in this section is complete from the mathematical viewpoint. Then, we may state the following.
Since the equation system that determines the generalized ${\tilde {\bar {\bf E}}}({\bf k}, t)$ and ${\tilde {\bar {\bf H}}}({\bf k}, t)$  fields is linear and
due to the average value of the field ${\tilde {\bar \rho}}({\bf k}, t)$ is zero, the thermal electromagnetic field is the random Gaussian field with zero average.

The initial conditions for calculation of mathematical expectations of various random functions of the ${\tilde {\bar {\bf  E}}} ({\bf k}, t)$, ${\tilde {\bar {\bf  H}}} ({\bf k}, t)$ and ${\tilde {\bar \rho}} ({\bf k}, t)$ fields become insignificant after the temporal period which is much longer than the time $\kappa {\bar L}^2/\varkappa$. (We also note that this temporal period should be much larger than the characteristic time $\tau$ associated with the thermal radiation, so that the value $\hbar \tau^{-1}$  should be of the average temperature order). Then, as the field ${\tilde {\boldsymbol\varphi}}({\bf x}, t)$ is stationary on $t$, so we may also consider the stochastic fields $\{{\tilde {\bar  {\bf E}}} ({\bf k}, t)$, ${\tilde {\bar {\bf H}}} ({\bf k}, t)\}$ which obey Eqs. (12)-(14) as stationary ones. Due to this, we neglect the dependence on time of the temperature distribution $T({\bf x}, t)$ in the amplitude $a({\bf x}, t ; T)$ and, consequently, in  sources ${\tilde {\bar {\bf j}}}({\bf k}, t)$,  ${\tilde \rho} ({\bf k}, t)$ when the transition to the asymptotic values are calculated. The such a disregard of temporal dependencies  corresponds to the transition in the asymptotic region $t  \gg \kappa {\bar L}^2/ \varkappa   \sim 10^{-7}$sec.

When we study the constructed stationary process, it is natural to pass from the evolution Eqs.(12)-(14) to the equations of spectral amplitudes of these fields. They are generalized functions on frequency $\omega$,
$${\tilde {\bar {\bf E}}}({\bf k}, t) = \int\limits^\infty_{- \infty} {\tilde {\boldsymbol {\cal E}}}({\bf k}, \omega) e^{i\omega t} d \omega\,, \quad
{\tilde {\bar {\bf H}}}({\bf k}, t) = \int\limits^\infty_{- \infty} {\tilde {\boldsymbol {\cal H}}}({\bf k}, \omega) e^{i\omega t} d \omega\,, \eqno (20) $$
$${\tilde {\bar {\bf j}}}({\bf k}, t) = \int\limits^\infty_{- \infty} {\tilde {\boldsymbol \iota}}({\bf k}, \omega) e^{i\omega t} d \omega\,, \quad
{\tilde {\bar {\bf \rho}}}({\bf k}, t) = \int\limits^\infty_{- \infty} {\tilde {\cal \varrho}}({\bf k}, \omega) e^{i\omega t} d \omega \eqno (21) $$
where the generalized random field ${\tilde \iota}_l({\bf k}, \omega)$ that defines the spectral expansion of fluctuating current density, is given by the formula
$${\tilde {\boldsymbol \iota}}({\bf k}, \omega) = \frac 1{(2\pi)^4}\int\limits_{{\Bbb R}^4}\exp\big(- i\omega t - i({\bf k}, {\bf x})\big)a({\bf x}, t; T ){\tilde
{\boldsymbol \varphi}}({\bf x}, t)d{\bf x}dt\,. \eqno (22)$$

Substituting these expansions into Eqs. (12-14) and using the uniqueness of Fourier's images, we obtain the following complete equations system:
$$i \omega{\tilde {\boldsymbol {\cal E}}}({\bf k}, \omega)\ + \gamma {\tilde {\boldsymbol {\cal E}}}({\bf k}, \omega) +  \frac {4 \pi}\varepsilon\, {\tilde {\boldsymbol \iota}}({\bf k}, \omega) = \frac {ic}{\varepsilon} [{\bf k}, {\tilde  {\boldsymbol {\cal H}}}({\bf k}, \omega)]  \,, \eqno  (23)$$
$${\tilde {\boldsymbol {\cal H}}}({\bf k}, \omega)\ = - \frac{c}{\mu \omega}[{\bf k}, {\tilde  {\boldsymbol {\cal E}}}({\bf k}, \omega)]\,, \quad ({\bf k}, {\tilde {\boldsymbol {\cal E}}}({\bf k}, \omega)) = - \frac {4\pi i} \varepsilon\, {\tilde \varrho}({\bf k}, \omega) \,,  \quad   ({\bf k}, {\tilde  {\boldsymbol {\cal H}}}({\bf k}, \omega)) = 0\,, \eqno (24)$$
$$i\omega {\tilde \varrho} ({\bf k}, \omega) + \gamma {\tilde \varrho}({\bf k}, \omega) + i ({\bf k}, {\tilde {\boldsymbol \iota}}({\bf k}, \omega)) = 0\,, \eqno (25)$$
Solutions of the system are performed by following formulas:
$${\tilde {\boldsymbol {\cal E}}}({\bf k}, \omega) =  i\frac {4 \pi} {\varepsilon} \cdot \frac {\Big((\omega^2 - i \omega \gamma){\tilde {\boldsymbol \iota}}({\bf k}, \omega)-
{\bar c}^2 ({\bf k}, {\tilde {\boldsymbol \iota}}({\bf k}, \omega)){\bf k} \Big)}{(\omega - i \gamma)(\omega^2 - {\bar c}^2 {\bf k}^2 - i \omega \gamma)}\,, \eqno (26)$$
$${\tilde {\boldsymbol {\cal H}}}({\bf k}, \omega) = - i\frac {4 \pi c} {\varepsilon \mu}\cdot \frac {[{\bf k},  {\tilde {\boldsymbol \iota}}({\bf k}, \omega)]}{(\omega^2 - {\bar c}^2 {\bf k}^2 -  i \omega \gamma)}\,.\eqno (27)$$
Here, Fourier's images ${\tilde {\boldsymbol {\cal E}}({\bf k}, \omega)}$, ${\tilde {\boldsymbol {\cal H}}({\bf k}, \omega)}$, ${\tilde {\boldsymbol \iota}}({\bf k}, \omega))$, ${\tilde \varrho}({\bf k}, \omega)$ are some generalized functions.
\smallskip

{\bf 5. Energy flux density at the stationary regime.} Let us calculate the average value of the energy flux density $S_j({\bf x}, t)$, $j = 1, 2, 3$, if
each its component is equal to
$$S_j({\bf x}, t)  = \frac c {4\pi}\avel [{\bf E}({\bf x}, t), {\bf H}({\bf x}, t)]_j\aver  = $$
$$ = \epsilon_{jll'} \frac c {4\pi}\int\limits_{{\Bbb R}^8} \exp\Big[i({\bf k} - {\bf k}', {\bf x}) + i (\omega - \omega')t\Big]\,\, \avel {\tilde {\cal E}}_l({\bf k}, \omega) {\tilde {\cal H}}_{l'}^*({\bf k}', \omega') \aver\,\, d {\bf k} d {\bf k}' d \omega d \omega' \eqno (28)$$
where $\epsilon_{jll'}$ is completely antisymmetric pseudotensor in ${\Bbb R}^3$ (the  Levi-Civita symbol).

The mathematical expectation in Eq.(28) is calculated on the basis of the explicit   expres\-sions (26) and (27) of the random fields ${\tilde {\boldsymbol {\cal E}}}({\bf k}, \omega), {\tilde {\boldsymbol {\cal H}}}({\bf k}, \omega)$. These fields are Gaussian, since the fluctuation current ${\tilde {\boldsymbol \iota}}({\bf x}, t)$ density is the random Gaussian field and due to linear transforma\-tions of them. Therefore, the mathematical expectation  is expressed by the correlation function of the field ${\tilde {\boldsymbol \iota}}({\bf k}, \omega)$,
$$\avel{\tilde {\cal E}}_l({\bf k}, \omega) {\tilde {\cal H}}_{l'}^*({\bf k}', \omega') \aver = $$
$$= - \frac {(4 \pi)^2 c} {\varepsilon^2 \mu} \cdot  \frac {\epsilon_{l'mm'} {k'}_m  \Big(\omega(\omega - i \gamma) \delta_{ln}  - {\bar c}^2 k_l k_{n}\Big)}{(\omega - i \gamma)(\omega^2 - {\bar c}^2 {\bf k}^2 - i \omega \gamma)(\omega'^2 - {\bar c}^2 {\bf k'}^2 +  i \omega' \gamma)}\,
\avel  {\tilde {\boldsymbol \iota}}_n({\bf k}, \omega){\tilde {\boldsymbol \iota}}^{\,*}_{m'}({\bf k'}, \omega')\aver\,, \eqno (29)$$
$$\avel  {\tilde {\boldsymbol \iota}}_l({\bf k}, \omega){\tilde {\boldsymbol \iota}}^{\,*}_{l'}({\bf k'}, \omega')\aver = \frac {\delta_{ll'}}{(2 \pi)^7} \int\limits_{{\Bbb R}^7} a ({\bf x}, t; T)
a ({\bf x}', t; T) K (|{\bf x} - {\bf x'}|)\exp\Big[i\Big(({\bf k}', {\bf x}') - ({\bf k}, {\bf x})\Big)\Big] d {\bf x}d {\bf x}' dt\,.  \eqno (30)$$

Substituting Eq.(28) into Eq.(29) and using the tensor identity $\epsilon_{jll'} \epsilon_{l'mn} = \delta_{jm}\delta_{ln} - \delta_{jn}\delta_{lm}$, we take into account  the expression (30) of the correlation function $\avel {\tilde {\boldsymbol \iota}}_n({\bf k}, \omega){\tilde {\boldsymbol \iota}}^*_{m'}({\bf k'}, \omega')\aver$
corresponding to isotropic stochastic fluctuations of the current density. As a result, we obtain the following expression of the energy flux density
$$S_j ({\bf x}, t) = \int\limits_{{\Bbb R}^7} R_j ({\bf x} - {\bf y}_1, t - s; {\bf x} - {\bf y}_2, t - s) K(|{\bf y}_1 - {\bf y}_2|) a({\bf y}_1, s; T)
a ({\bf y}_2, s; T) d {\bf y}_1 d {\bf y}_2 ds\,, \eqno (31)$$
where
$$R_j ({\bf x}, t; {\bf x}', t')= \frac {1}{(2\pi)^8} \int\limits_{{\Bbb R}^8} {\bar R}_j ({\bf k}, \omega; {\bf k}', \omega')
\exp\Big[i\big(({\bf k}, {\bf x}) - ({\bf k}', {\bf x}')\big) + i (\omega t - \omega' t')\Big] d {\bf k} d {\bf k}' d \omega d \omega'\,, \eqno (32)$$
$${\bar R}_j ({\bf k}, \omega; {\bf k}', \omega') = - R \frac{\Big(k'_j(2\omega (\omega - i \gamma) - {\bar c}^2 {\bf k}^2) + {\bar c}^2 k_j (k_m k'_m)\Big)} {(\omega - i \gamma)(\omega^2 - {\bar c}^2 {\bf k}^2 - i \omega \gamma)(\omega'^2 - {\bar c}^2 {\bf k'}^2 +  i \omega' \gamma)}\,,  \eqno (33)$$
$$R = {4 \pi {\bar c}^{\,2}}/ {\varepsilon}\,. \eqno (34)$$

From formulas (32) and (33), we have the function
$$R_j ({\bf x}, t; {\bf x}', t')= - \frac {R}{(2\pi)^8} \int\limits_{{\Bbb R}^8}
\frac{\Big(k'_j(2\omega (\omega - i \gamma) - {\bar c}^2 {\bf k}^2) + {\bar c}^2 k_j (k_m k'_m)\Big)} {(\omega - i \gamma)(\omega^2 - {\bar c}^2 {\bf k}^2 - i \omega \gamma)(\omega'^2 - {\bar c}^2 {\bf k'}^2 +  i \omega' \gamma)} \times $$
$$ \times
\exp\Big[i\big(({\bf k}, {\bf x}) - ({\bf k}', {\bf x}')\big) + i (\omega t - \omega' t')\Big] d {\bf k} d {\bf k}' d \omega d \omega'\,, $$
that defines contributions of two radiation sources into the energy flux density at the point ${\bf x}$ (since the energy flux density is proportional to the square of electromagnetic field). These sources are in different spatial points with ${\bf y}_1$ and ${\bf y}_2$ radius-vectors. So, the function  $R_j ({\bf x}, t; {\bf x}', t')$ should  be represented in the form
$$R_j ({\bf x}, t; {\bf x}', t') = - R\Big[iU({\bf x}, t) \nabla_j' V^*({\bf x}', t') + \phantom{AAAAAAAAAAAAAA}$$
$$ \phantom{AAAAAAAAAAAAAA} + \dot V({\bf x}, t) \nabla_j' V^*({\bf x}', t') - i{\bar c}^2 \nabla_m \nabla_j W({\bf x}, t)\nabla_m' V^*({\bf x}', t')\Big] \eqno (35)$$
where the operators $\nabla_j$ and $\nabla_m'$, $j, m' = 1, 2, 3$ denote gradients on vectors ${\bf x}$ and ${\bf x}'$, respectively, and the dot denotes the differentiation on $t$.
The scalar fields $U({\bf x}, t)$, $V({\bf x}, t)$, $W({\bf x}, t)$ are given by the following integral representations which are some generalized functions corresponding:
$$U({\bf x},t)= \frac {1}{(2\pi)^4}\int\limits_{{\Bbb R}^4}\frac {\exp(i({\bf k},{\bf x}) + i \omega t)} {\omega - i \gamma}d {\bf k}d \omega\,, \eqno (36)$$
$$V({\bf x},t)= \frac {1}{(2\pi)^4}\int\limits_{{\Bbb R}^4}\frac {\exp(i({\bf k},{\bf x}) + i \omega t)} {\omega^2 - {\bar c}^2 {\bf k}^2 - i \omega \gamma}d {\bf k}d \omega\,, \eqno (37)$$
$$W({\bf x},t)= \frac {1}{(2\pi)^4}\int\limits_{{\Bbb R}^4}\frac {\exp(i({\bf k},{\bf x}) + i \omega t)} {(\omega - i \gamma)(\omega^2 - {\bar c}^2 {\bf k}^2 - i \omega \gamma)}d {\bf k}d \omega\,. \eqno (38)$$
Besides, functions $U({\bf x}, t)$ and $W ({\bf x}, t)$ have purely imaginary values, and the function $V ({\bf x}, t)$ has real values.

In accordance with Eq.(35), the flux density $S_j((\bf x),t)$ breaks to three parts:
$$S_j ({\bf x}, t) = S_j^{(u)} ({\bf x}, t) + S_j^{(v)} ({\bf x}, t) + S_j^{(w)} ({\bf x}, t) \eqno (39)$$
where, according to Eq.(31) and Eq.(35), each term of $S_j((\bf x),t)$ has the form:
$$S_j^{(u)} ({\bf x}, t) = - iR\int_{{\Bbb R}^7} U({\bf x} - {\bf y}_1, t - s) \nabla_j V^*({\bf x} - {\bf y}_2, t - s) \times \phantom{AAAAAAAAAAAAAAA} $$
$$ \phantom{AAAAAAAAAAAAAAAAAAAA} \times
K(|{\bf y}_1 - {\bf y}_2|) a({\bf y}_1, s; T)a ({\bf y}_2, s; T) d {\bf y}_1 d {\bf y}_2 ds\,, \eqno (40)$$
$$S_j^{(v)} ({\bf x}, t) = - R\int_{{\Bbb R}^7}[\dot V({\bf x} - {\bf y}_1, t -s)] [\nabla_j V^*({\bf x} - {\bf y}_2, t - s)]\times \phantom{AAAAAAAAAAAAAAA} $$
$$ \phantom{AAAAAAAAAAAAAAAAAAAA} \times
K(|{\bf y}_1 - {\bf y}_2|) a({\bf y}_1, s; T)a ({\bf y}_2, s; T) d {\bf y}_1 d {\bf y}_2 ds\,, \eqno (41)$$
$$S_j^{(w)} ({\bf x}, t) = i{\bar c}^2  R\int_{{\Bbb R}^7}[\nabla_m \nabla_j W({\bf x} - {\bf y}_1, t - s)][\nabla_m V^*({\bf x} - {\bf y}_2, t -s)]\times \phantom{AAAAAAAAAA} $$
$$ \phantom{AAAAAAAAAAAAAAAAAAAA} \times K(|{\bf y}_1 - {\bf y}_2|) a({\bf y}_1, s; T)a ({\bf y}_2, s; T) d {\bf y}_1 d {\bf y}_2 ds\,, \eqno (42)$$
\smallskip

{\bf 6. Asymptotic expressions of generalized functions $U, V, W$.} In this section we find asymptotic formulas of generalized functions $U({\bf x}, t)$, $V({\bf x}, t)$, $W({\bf x}, t)$ which determine the contributions $S_j^{(p)} ({\bf x},t)$, $p \in \{u, v, w\}$ to the flux density $S_j ({\bf x}, t)$ when the small parameter\ \ $\gamma L/{\bar c} \to 0$. On the one hand, such  a procedure is necessary due to the fact that functions $V({\bf x}, t)$, $W({\bf x}, t)$ are not calculated exactly in terms of standard generalized functions. It makes very complicated formulas (41) and (42). On the other hand, we must to calculate the value $S_j ({\bf x}, t)$ so that it may be  used for application  in the theory of radiation  heat transfer in semitransparent semiconductor crystals.

For the generalized function $U$, one can find easily the explicit form
$$U({\bf x},t)=  \frac {\delta({\bf x})}{2\pi} \int\limits_{\Bbb R} \frac {e^{i \omega t}} {\omega - i \gamma}d \omega
= i \Theta(t) \delta({\bf x})e^{- \gamma t}  \eqno (43)  $$
where we have used the integral representation of three-dimensional $\delta$-function
$$\delta ({\bf x})= \frac {1}{(2\pi)^3}\int\limits_{{\Bbb R}^{\,3}}\exp(i({\bf k},{\bf x})) d {\bf k} $$
and the integral representation of the Heaviside $\Theta(\cdot)$-function.

Functions $V({\bf x}, t)$ and $W({\bf x}, t)$ do not have such  a simple explicit representation. Therefore, we find their asymptotic representations when the parameter $\gamma L/{\bar c}$ tends to zero. As a result, we have found the following asymptotic formulas at $r > 0$, $t > 0$:
$$V({\bf x}, t)= - \frac {\Theta(t)}{4\pi {\bar c} r}\, e^{-\gamma t/2}\delta(r - {\bar c}t)\,,\eqno (44)$$
$$W({\bf x}, t)  \sim - \frac{i \Theta (t)}{4 \pi{\bar c}^2} \Big[\frac {e^{- \gamma t}}r - \phantom{AAAAAAAAAAAAAAAAAAAAAAAAA}$$
$$\phantom {AAA} - \frac 1{2r}\, e^{- \gamma t/2}\,
\Big({\rm sgn}(r+ {\bar c}t) + {\rm sgn}(r- {\bar c}t)  -  \frac \gamma {2{\bar c}}\big[ |r + {\bar c}t| - |r- {\bar c}t|\big]\Big) -  \frac {2 {\dot g}(t)}{\pi {\bar c}}\Big] \eqno (45)$$
where
$$g (t) \equiv e^{- \gamma t/2}\int\limits_0^{\pi/2} {\rm sh}\Big(\frac {\gamma t}2 \cos \eta\Big){d\eta} = \int\limits_0^{\pi/4}\Big(\exp\big( - \gamma t \sin^2 \eta\big) - \exp\big( - \gamma t \cos^2 \eta\big) \Big){d\eta}\,. \eqno (46)$$
\smallskip

{\bf 7. Integral representations $S_j^{(u)} ({\bf x}, t)$, $S_j^{(v)} ({\bf x}, t)$, $S_j^{(w)} ({\bf x}, t)$.} Now, we obtain such asymptotic integral representations of functions  $S_j^{(u)} ({\bf x}, t)$, $S_j^{(v)} ({\bf x}, t)$, $S_j^{(w)} ({\bf x}, t)$ which do not contain  $\delta$-function singularities when
limit transitions $\gamma L/{\bar c} \to 0$ and  $\varkappa/ {L {\bar c} \kappa} \to 0$ are made.

Further, we denote gradients on variables ${\bf y}_j$ by $\nabla^{(j)}$, $j = 1,2$ respectively.
Since all subintegral expressions of integrals in Eqs.(40-42)  contain  the gradient of function $V({\bf x}, t)$ with $\delta$-function singularity,
we fulfill integrations on the variable ${\bf y}_2$ in them  by parts. Then,  we calculate the integral on $s$ by means of the $\delta$-function.  We use also
the asymptotic formula (44). As a result of transformations pointed out, we obtain the following formulas:

$$S_j^{(u)} ({\bf x}, t) = - i\frac R{4\pi  {\bar c}^2}  \int\limits_{{\Bbb R}^6}\frac {e^{-\gamma |{\bf y}_2|/2{\bar c}}}{|\,{\bf y}_2|} \nabla^{(2)}_j\Big[K(|{\bf y}_2 - {\bf y}_1|)a({\bf x} - {\bf y}_1, t - s; T)a({\bf x} - {\bf y}_2, t - s; T)\Big]_{s = {|{\bf y}_2|}/{{\bar c}}}\times $$
$$\times U({\bf y}_1, |{\bf y}_2|/{\bar c}) d {\bf y}_1 d {\bf y}_2\,, \eqno (47)$$
$$S_j^{(v)} ({\bf x}, t)  = - \frac R{4\pi  {\bar c}^2}   \int\limits_{{\Bbb R}^6}\frac{e^{-\gamma |{\bf y}_2|/2{\bar c}}}{|\,{\bf y}_2|}\nabla_j^{(2)}\Big[
K(|{\bf y}_2 - {\bf y}_1|)a({\bf x} - {\bf y}_1, t - s; T)a({\bf x} - {\bf y}_2, t - s; T)\Big]_{s = {|{\bf y}_2|}/{{\bar c}}}\times $$
$$\times \dot V({\bf y}_1, |{\bf y}_2|/{\bar c})d {\bf y}_1 d {\bf y}_2\,, \eqno (48)$$
$$S_j^{(w)} ({\bf x}, t)  = \frac {iR}{4\pi}\int\limits_{{\Bbb R}^6}\frac {e^{-\gamma |{\bf y}_2|/2{\bar c}}}{|\,{\bf y}_2|}\nabla^{(1)}_j \nabla_m^{(1)}\nabla^{(2)}_m\Big[K(|{\bf y}_2 - {\bf y}_1|)a({\bf x} - {\bf y}_1, t - s; T)a({\bf x} - {\bf y}_2, t - s; T)\Big]_{s = {|{\bf y}_2|}/{{\bar c}}}$$
$$\times\ W({\bf y}_1, |{\bf y}_2|/{\bar c}) d {\bf y}_1 d {\bf y}_2\,. \eqno (49) $$

To obtain asymptotic  expressions of these integrals at the limit $\varkappa/ {L {\bar c} \kappa} \to 0$, we use that the temperature distribution $T({\bf x}, t)$ changes very slowly on spatial coordinates in comparison with changing of correlation function $K(\cdot)$. Therefore, we neglect results of  operators $\nabla^{(1)}$ and $\nabla^{(2)}$ actions on amplitudes    $a({\bf x} - {\bf y}_1, t; T)$ and $a({\bf x} - {\bf y}_2, t; T)$, respectively:
$$S_j^{(u)} ({\bf x}, t) = -\frac {iR}{4\pi  {\bar c}^2}\,  \int\limits_{{\Bbb R}^6}\frac {e^{-\gamma |{\bf y}_2|/2{\bar c}}}{|\,{\bf y}_2|}\, U({\bf y}_1, |{\bf y}_2|/{\bar c})\times $$
$$\times \Big(\nabla^{(2)}_j K(|{\bf y}_2 - {\bf y}_1|)\Big) a({\bf x} - {\bf y}_1, t - {|{\bf y}_2|}/{{\bar c}}; T)a({\bf x} - {\bf y}_2, t - {|{\bf y}_2|}/{{\bar c}}; T) d {\bf y}_1 d {\bf y}_2\,, \eqno (50) $$
$$S_j^{(v)} ({\bf x}, t)  = -\frac R{4\pi  {\bar c}^2}   \int\limits_{{\Bbb R}^6}\frac{e^{-\gamma |{\bf y}_2|/2{\bar c}}}{|\,{\bf y}_2|}\,{\dot V}({\bf y}_1, |{\bf y}_2|/{\bar c})\times $$
$$\times \Big(\nabla_j^{(2)}K(|{\bf y}_2 - {\bf y}_1|)\Big) a({\bf x} - {\bf y}_1, t - {|{\bf y}_2|}/{{\bar c}}; T)a({\bf x} - {\bf y}_2, t - {|{\bf y}_2|}/{{\bar c}}; T)
d {\bf y}_1 d {\bf y}_2\,, \eqno (51)$$
$$S_j^{(w)} ({\bf x}, t)  = \frac {iR}{4\pi}\int\limits_{{\Bbb R}^6} \frac {e^{-\gamma |{\bf y}_2|/2{\bar c}}}{|\,{\bf y}_2|}\, W({\bf y}_1, |{\bf y}_2|/{\bar c})\times $$
$$\times  \Big(\nabla^{(1)}_j \nabla_m^{(1)} \nabla^{(2)}_m K(|{\bf y}_2 - {\bf y}_1|)\Big) a({\bf x} - {\bf y}_1, t - {|{\bf y}_2|}/{{\bar c}}; T)a({\bf x} - {\bf y}_2, t - {|{\bf y}_2|}/{{\bar c}}; T )d {\bf y}_1 d {\bf y}_2\,, \eqno (52)$$
\smallskip

{\bf 8. Asymptotic expression of the function $S_j ({\bf x}, t)$ at $r_0/L \to 0$.} Here, we calculate asymptotic expression of the energy flux density $S_j ({\bf x}, t)$ when  the correlation radius  $r_0/L$ tends to 0.

To do such a limit transition, it is necessary to introduce  explicitly  the parameter $r_0$ into these expressions. It is done by the following form of the correlation function  $K(|{\bf x}|) = r_0^{-3} Q({\bf x}^2/2r^2_0)$ that provides an independence of the integral  $\int K(|{\bf z}|)d {\bf z}$ on $r_0$. Since the subintegral function has singularities, the transition to the limit $r_0/L \to 0$ is not possible by means of the change of $K(|{\bf z}|)$ to  $K \delta ({\bf z})$ with a positive  constant $K$ in the subintegral expression.

After introduction the explicit dependence  of correlation function on $r_0$ into subintegral expressions in Eqs.(50-52), we make the following changes of integration variables ${\bf y}_1/r_0 \Rightarrow {\bf y}_1$ and ${\bf y}_2/r_0 \Rightarrow {\bf y}_2$. Then, passing to the limit  $r_0 \to 0$, we calculate main terms of asymptotic expansion of densities  $S_j^{(u)} ({\bf x}, t)$,  $S_j^{(v)} ({\bf x}, t)$, $S_j^{(w)} ({\bf x}, t)$, substituting the asymptotic expressions of functions $U({\bf x}, t)$, $V({\bf x}, t)$, $W({\bf x}, t)$ in corresponding integral. As a result, we obtain the following asymptotic formulas
$$S_j^{(u)} = a^2({\bf x}, t;T) \Big(\frac {r_0^{-2}R}{4\pi  {\bar c}^2}\Big)\int\limits_{{\Bbb R}^3}\frac {\nabla_j Q({\bf y}^2/2)}{|{\bf y}|}\,d {\bf y} + r_0^{-2}o(1) \eqno (53)$$
where the integral is equal to zero due to the spherical symmetry of the correlation function. Thus, we get finally $S_j^{(u)}= r_0^{-2}o(1)$ when $r_0 \to 0$.

Further, for the function  $S_j^{(v)} ({\bf x}, t)$ we obtain
$$S_j^{(v)} ({\bf x}, t)  = -\frac {R}{(4\pi  {\bar c})^2}   \int\limits_{{\Bbb R}^6}\frac{e^{-\gamma r_0|{\bf y}_2|/{\bar c}}}{|\,{\bf y}_2||{\bf y}_1|}\, \Big(\nabla_j^{(2)}Q(|{\bf y}_2 - {\bf y}_1|^2/2)\Big)\, \times \phantom{AAAAAAAAAA}$$
$$\Big(\frac \gamma {2{\bar c}} \,\delta(r_0(|{\bf y}_1| -  |{\bf y}_2|)) + \delta' (r_0(|{\bf y}_1| - |{\bf y}_2|))\Big)
a({\bf x} - r_0{\bf y}_1, t; T)a({\bf x} - r_0{\bf y}_2, t; T) d {\bf y}_1 d {\bf y}_2 \eqno (54) $$
where we have taken into account that $\delta$-function $\delta (|{\bf y}_2| + |{\bf y}_1|)$ does not give a contribution into the integral.

After the introduction of dimensionless integration variables, the first term is proportional to the small parameter $\gamma L/{\bar c}$.
However, it is necessary to ascertain that the  corresponding integral  has the order of $r^{-1}_0$ and we may neglect this term.
Using spherical coordinates, this integral is obtained in the form
$$\int\limits_{{\Bbb R}^6}\frac{e^{-\gamma r_0|{\bf y}_2|/{\bar c}}}{|{\bf y}_2||{\bf y}_1|}a({\bf x} - r_0{\bf y}_1, t)a({\bf x} - r_0{\bf y}_2, t) \Big[\nabla_j^{(2)}Q(({\bf y}_2 - {\bf y}_1)^2/2)\Big]\,\delta(r_0(|{\bf y}_1| -  |{\bf y}_2|))d {\bf y}_1 d {\bf y}_2 =$$
$$= r_0^{-1}\int\limits_0^\infty  \xi^3 \exp\Big[-\gamma r_0\xi/{\bar c}\Big]d \xi \times $$
$$\times \int\limits_{{\bf n}_1^2 = {\bf n}_2^2 = 1}    a({\bf x} - r_0\xi{\bf n}_1, t; T)a({\bf x} - r_0\xi{\bf n}_2, t;T) ({\bf n}_1 - {\bf n}_2)_jQ' (\xi^2|{\bf n}_2 - {\bf n}_1|^2/2)d \Omega({\bf n}_1)d \Omega({\bf n}_2)$$
where we introduce  integrations on $d\Omega({\bf n}_1)$ and $d \Omega ({\bf n}_2)$ that means integrations on unit spheres of vectors ${\bf y}_1$ and ${\bf y}_2$ respectively. The last integral is equal to zero, since the subintegral expression of its internal integral is antisymmetric comparatively to the permuta\-tion ${\bf n}_1 \Leftrightarrow {\bf n}_2$. Consequently,  we may really neglect the term with $\delta$-function.

Now, we calculate the contribution into asymptotic expression of the function $S_j^{(v)}({\bf x}, t)$ connected with the integral that contains $\delta'(\cdot)$-function.
To calculate this contribution,  it is convenient to return to the integral variable $r_0{\bf y}_j \Rightarrow {\bf y}_j$, $j = 1,2$, ${\bar R} = {R, r_0^{-3}}/{(4\pi {\bar c})^2}$,
$$- {\bar R} \int\limits_{{\Bbb R}^6}\frac{e^{-\gamma |{\bf y}_2|/{\bar c}}}{|{\bf y}_2||{\bf y}_1|}\,a({\bf x} - {\bf y}_1, t; T)a({\bf x} - {\bf y}_2, t; T)\Big[\nabla_j^{(2)}Q(({\bf y}_2 - {\bf y}_1)^2/2r_0^2)\Big]\delta' (|{\bf y}_1| - |{\bf y}_2|)d {\bf y}_1 d {\bf y}_2 = $$
$$ = - r^{-2}_0 {\bar R} \int\limits_{{\Bbb R}^3}\frac{e^{-\gamma |\,{\bf y}_2|/{\bar c}}}{|\,{\bf y}_2|}\, a({\bf x} - {\bf y}_2, t; T)\Big(C_{\parallel}({\bf y}_2){\bf n} + {\bf C}_\perp ({\bf y}_2)\Big)d {\bf y}_2\,, \eqno (55)$$
where  ${\bf n}  = {\bf y}_{2}/|{\bf y}_2|$ and
$$C_\parallel({\bf y}_2) = \int\limits_{{\Bbb R}^3} \frac {|\,{\bf y}_{2}| - ({\bf n}, {\bf y}_{1})}{|\,{\bf y}_1|}\, Q'((\,{\bf y}_2 - {\bf y}_1)^2/2r_0^2)\delta' (|\,{\bf y}_1| - |\,{\bf y}_2|)a({\bf x} - {\bf y}_1, t; T)d {\bf y}_1\,, $$
$${\bf C}_\perp ({\bf y}_2) =  - \int\limits_{{\Bbb R}^3}
\frac {{\bf y}_{1} - ({\bf y}_1, {\bf n}){\bf n}}{|\,{\bf y}_1|}\, Q'((\,{\bf y}_2 - {\bf y}_1)^2/2r_0^2)\delta' (|\,{\bf y}_1| - |{\bf y}_2|)a({\bf x} - {\bf y}_1, t; T)d {\bf y}_1$$
so that  $({\bf C}_\perp, {\bf n}) = 0$.

Passing to the limit $r_0 \to 0$, we find the limiting expression of introduced integrals
$$C_\parallel({\bf y}_2) = - \frac {2\pi Q_0}{|{\bf y}_{2}|} r_0^2 a ({\bf x} - {\bf y}_2, t; T) (1 + O(r_0^2)) \eqno (56)$$
where it is assumed that $Q_0 = Q(0) < \infty$. Similarly, we  find that  ${\bf C}_\perp ({\bf y}_2) =  o(r_0^3)$.

Substituting asymptotic expressions $C_\parallel ({\bf y}_2)$ and ${\bf C}_\perp ({\bf y}_2)$ into Eq.(55) and after that into Eq.(54), we find the final asymptotic formula for $ S_j ^ {(v)} ({\bf x}, t) $,
$$S^{(v)}_j ({\bf x}, t) =  \frac {r_0^{-3}R Q_0}{8\pi  {\bar c}^2} \int\limits_{{\Bbb R}^3}\frac{y_je^{-\gamma |{\bf y}|/{\bar c}}}{|{\bf y}|^3}\, a^2({\bf x} - {\bf y}, t; T)d {\bf y}\,. \eqno (57)$$
The obtained formula shows that we may neglect the function $S^{(u)}_j ({\bf x}, t)$ when the main term of the asymptotic flux density $S_j ({\bf x}, t)$ is calculated.

Now, we pass to calculation of asymptotic formula of $S_j^{(w)} ({\bf x}, t)$. Substituting the asympto\-tical expression (45) of the function $W({\bf x}, t)$ into Eq.(52) and producing  the replacements of the integration variables $r_0{\bf y}_j \Rightarrow {\bf y}_j$, $j = 1,2$, we find

$$S_j^{(w)} ({\bf x}, t)  = - \frac {r_0^{-3}R}{(4\pi{\bar c})^2}\int\limits_{{\Bbb R}^6}\frac {e^{-\gamma |{\bf y}_2|/{\bar c}}}{|{\bf y}_2||{\bf y}_1|}\Big(\nabla^{(1)}_j \Delta^{(1)} Q(({\bf y}_2 - {\bf y}_1)^2/2r_0^2)\Big)a({\bf x} - {\bf y}_1, t; T)a({\bf x} - {\bf y}_2, t; T) \times $$
$$\times  \Big[e^{- \gamma |{\bf y}_2|/2{\bar c}} - \frac 12\, \Big(1 + {\rm sgn}(|{\bf y}_1| - |{\bf y}_2|)\Big)\Big]d {\bf y}_1 d {\bf y}_2 \equiv S_j^{(w, 1)}({\bf x}, t) + S_j^{(w, 2)}({\bf x}, t)  \eqno (58)$$
where we have neglected the term connected with the function   ${2 {\dot g}(t)}$ because of its smallness.

Let us consider the first summand. We replace the integration variables according to formulas ${\bf y} = ({\bf y}_1 + {\bf y}_2)/2$, ${\bf z} = {\bf y}_1 - {\bf y}_2$ which have the jacobian equal to 1. After that we replace ${\bf z}/r_0 \Rightarrow {\bf z}$. As a result, we obtain
$$S_j^{(w, 1)}({\bf x}, t) \equiv - \frac {r_0^{-3}R}{(4\pi{\bar c})^2}\int\limits_{{\Bbb R}^6}\frac {\exp\big[-3\gamma |{\bf y} - r_0 {\bf z}/2|/2{\bar c}\big]}{|{\bf y} - r_0{\bf z}/2||{\bf y} + r_0 {\bf z}/2|}\,\Big[\nabla^{({\bf z})}_j \Delta^{({\bf z})} Q({\bf z}^2/2)\Big] \times \phantom{AAAAAA}$$
$$\phantom{AAAAAAAAAAAAAAAAA} \times a({\bf x} - {\bf y} - r_0{\bf z}/2, t; T)a({\bf x} - {\bf y} + r_0 {\bf z}/2, t; T) d {\bf y} d {\bf z}\,.$$

From here, passing to the limit $r_0 \to 0$, we find that the asymptotic term of the function $S_j^{(w, 1)}({\bf x}, t)$ is proportional to $r_0^{-3}$, is equal to zero, since  it is equal to zero the limiting expression of the following integral,
$$\int\limits_{{\Bbb R}^3}\frac {e^{-3\gamma |{\bf y}|/2{\bar c}}}{{\bf y}^2}\,a^2({\bf x} - {\bf y}, t; T) d {\bf y}\int\limits_{{\Bbb R}^3}
\Big[\nabla^{({\bf z})}_j \Delta^{({\bf z})} Q({\bf z}^2/2)\Big]d {\bf z} = 0\,, $$
due to the fact that replacement of the integration variable ${\bf z} \Rightarrow - {\bf z}$ in the subintegral expression of last integral changes its sign to the opposite one.

At last, we consider the second term in Eq.(58). Taking into account that $[1 + {\rm sgn}(|{\bf y}_1| - |{\bf y}_2|)]/2 = \theta (|{\bf y}_1| - |{\bf y}_2|)$, we have
$$S_j^{(w, 2)}({\bf x}, t) \equiv \frac {r_0^{-3}R}{(4\pi{\bar c})^2}\int\limits_{{\Bbb R}^6: |{\bf y}_1| > |{\bf y}_2|}\frac {e^{-\gamma |{\bf y}_2|/{\bar c}}}{|{\bf y}_2||{\bf y}_1|}\Big[\nabla^{(1)}_j\Delta^{(1)} Q(({\bf y}_2 - {\bf y}_1)^2/2r_0^2)\Big]\times \phantom{AAAAAAAAA}$$
$$\phantom{AAAAAAAAAAAAAAAAAAAA} \times a({\bf x} - {\bf y}_1, t; T)a({\bf x} - {\bf y}_2, t; T)d {\bf y}_1 d {\bf y}_2\,.$$

Producing replacements of integration variables which are analogous to  ones produced at the analysis of the function $S_j^{(w, 1)}({\bf x}, t)$, we obtain the following expression:
$$S_j^{(w, 2)}({\bf x}, t) \equiv  \frac {r_0^{-3}R}{(4\pi{\bar c})^2}\int\limits_{{\Bbb R}^6: ({\bf y}, {\bf z}) > 0}\frac {\exp\big[-\gamma |{\bf y} - r_0 {\bf z}/2|/{\bar c}\big]}{|{\bf y} - r_0{\bf z}/2||{\bf y} + r_0 {\bf z}/2|}\,\Big[\nabla^{({\bf z})}_j \Delta^{({\bf z})} Q({\bf z}^2/2)\Big] \times \phantom{AAAAAA}$$
$$\phantom{AAAAAAAAAAAAAAAAA} \times a({\bf x} - {\bf y} - r_0{\bf z}/2, t; T)a({\bf x} - {\bf y} + r_0 {\bf z}/2, t; T) d {\bf y} d {\bf z}\,.$$
Further, passing to the limit $r_0 \to 0$, we obtain the main asymptotic term of analyzed function in the following form:
$$S_j^{(w, 2)}({\bf x}, t) \equiv \frac {r_0^{-3}R}{(4\pi{\bar c})^2}\int\limits_{{\Bbb R}^3}\frac {e^{-\gamma |{\bf y}|/{\bar c}}}{{\bf y}^2}\,a^2({\bf x} - {\bf y}, t; T) d {\bf y}\int\limits_{{\Bbb R}^3:({\bf y}, {\bf z}) > 0}
\Big[\nabla^{({\bf z})}_j \Delta^{({\bf z})} Q({\bf z}^2/2)\Big]d {\bf z}\,. \eqno (59)$$
We transform the internal integral, using the formula of Gaussian type
$$\int\limits_{{\Bbb R}^3:({\bf y}, {\bf z}) > 0}
\Big[\nabla^{({\bf z})}_j \Delta^{({\bf z})} Q({\bf z}^2/2)\Big]d {\bf z} = - {\bf n}_j\int\limits_{{\Bbb R}^2}
\Delta^{({\bf z})} Q({\bf z}^2/2) d \Sigma ({\bf z}) \eqno (60)$$
where ${\bf n} = {\bf y}/|{\bf y}|$ and the integration is fulfilled over the plane that is perpendicular to ${\bf n}$ and contains the point ${\bf z} = 0$. It is assumed that $\Delta^{({\bf z})} Q({\bf z}^2/2)$ tends by sufficiently rapid way to zero in ${\Bbb R}^3$ when $|{\bf z}| \to \infty$.
Using polar coordinates $\langle \eta, \alpha \rangle$ on the integration plane in the last integral, we obtain the following result:
$$\int\limits_{{\Bbb R}^2} \Delta^{({\bf z})} Q({\bf z}^2/2) d \Sigma ({\bf z}) = 2\pi \int\limits^\infty_0\eta\, Q'(\eta^2/2)d \eta = - 2\pi Q_0\,.$$
Substituting this expression into Eq.(59) and after that into Eq.(58), we take into account that the coefficient at $r_0^{-3}$ in the asymptotic expression of $S_j^{(w, 1)} ({\bf x}, t)$ vanishes. Then, we obtain the asymptotic expression of the function:
$$S_j^{(w)} ({\bf x}, t) =   \frac {r_0^{-3}R Q_0}{8\pi  {\bar c}^2} \int\limits_{{\Bbb R}^3}\frac{y_je^{-\gamma |{\bf y}|/{\bar c}}}{|{\bf y}|^3}\, a^2({\bf x} - {\bf y}, t; T)d {\bf y}\,. \eqno (61)$$

The sum of the expression (61) and the main asymptotic term of the function $S_j^{(v)}({\bf x}, t)$ (57) which have the same order of $r_0 \to 0$, we find the final asymptotic expression of the energy flux density of fluctuating electromagnetic field
$$S_j ({\bf x}, t) =   \frac {r_0^{-3}R Q_0}{4\pi  {\bar c}^2} \int\limits_{{\Bbb R}^3}\frac{y_je^{-\gamma |{\bf y}|/{\bar c}}}{|{\bf y}|^3}\, a^2({\bf x} - {\bf y}, t; T)d {\bf y}\,.   \eqno (62)$$

{\bf 9. Conclusion.} From the result obtained in the previous section, we find  the divergence $(\nabla, {\bf S})$ of the energy flux density. It leads us to the main result of our work. We get the following evolution equation of thermal transfer with the account of the heat radiation conduction:
$$\kappa(T) {\dot T}({\bf x}, t) = (\nabla, \varkappa(T) \nabla) T({\bf x}, t) + \phantom{AAAAAAAAAAAAAAAA} $$
$$\phantom{AA} + K\Big({\gamma} {{\bar c}}^{-1}\,\int\limits_{{\Bbb R}^3} \frac{\exp\Big\{- \gamma |{\bf x} - {\bf y}|/{\bar c}\Big\}}{|{\bf x} - {\bf y}|^2}a^2({\bf y}, t; T)d {\bf y} - 4 \pi a^2 ({\bf x}, t; T) \Big)\,, \eqno (63)$$
where we have denoted $K =   {r_0^{-3}R Q_0}/{4 \pi {\bar c}^2}$.

The derived equation may be obtained, in principal,  on the basis of arguments used in the theory of radiation transfer in medium (see \cite{Spar}-\cite{Petr}). At the same time, it should be noted that there is the significant difference from the formula of standard theory. If we apply radiation transfer theory to the physical problem under study and the energy flux density is derived on the basis of geometric optics presentation,  then, the obtained formula differs from Eq.(63) by the supplement multiply $|{\bf x} - {\bf y}|^{-2}$ in integral kernel. Appearance of this multiply is associated with the presence of isotropic <<dissipation>> of radiation that is emitted by each spatial point of the medium.

\end{document}